\documentclass[10pt,twocolumn,letterpaper]{article}
\usepackage[utf8]{inputenc}
\usepackage{times}
\usepackage{epsfig}
\usepackage{graphicx}
\usepackage{amsmath}
\usepackage{amssymb,caption}
\usepackage[normalem]{ulem}
\usepackage[pagebackref=true,breaklinks=true,colorlinks,bookmarks=false]{hyperref}
\usepackage{authblk}
\date{}
\title{Scanner Invariant Multiple Sclerosis Lesion Segmentation from MRI}

\author[1,6]{Shahab Aslani}
\author[1,4]{Vittorio Murino}
\author[2]{Michael Dayan}
\author[5]{Roger Tam}
\author[1,3]{Diego Sona}
\author[6]{Ghassan Hamarneh}

\affil[1]{Pattern Analysis and Computer Vision, Istituto Italiano di Tecnologia, Genoa, Italy}
\affil[2]{Human Neuroscience Platform, Fondation Campus Biotech, Geneva, Switzerland}
\affil[3]{NeuroInformatics Laboratory, Fondazione Bruno Kessler, Trento, Italy}
\affil[4]{Computer Science Department, University of Verona, Italy}
\affil[5]{Department of Radiology and School of Biomedical Engineering, University of British Columbia, Vancouver, Canada}
\affil[6]{School of Computing Science, Simon Fraser University, Vancouver, Canada}

\begin{document}
\maketitle

\section{Abstract}
This paper presents a simple and effective generalization method for magnetic resonance imaging (MRI) segmentation when data is collected from multiple MRI scanning sites and as a consequence is affected by (site-)domain shifts. We propose to integrate a traditional encoder-decoder network with a regularization network. This added network includes an auxiliary loss term which is responsible for the reduction of the domain shift problem and for the resulting improved generalization. The proposed method was evaluated on multiple sclerosis lesion segmentation from MRI data. We tested the proposed model on an in-house clinical dataset including 117 patients from 56 different scanning sites. In the experiments, our method showed better generalization performance than other baseline networks.

\begin{keywords}
Magnetic Resonance Imaging, Multiple Sclerosis, Lesion Segmentation, Domain Generalization, 
\end{keywords}

\section{Introduction}
\label{sec:intro}
Deep learning  models, in particular Convolutional Neural Networks (CNNs) \cite{lecun1998gradient} have shown excellent performance in a large variety of computer vision tasks, including image classification \cite{huang2017densely}, object detection \cite{girshick2014rich}, semantic segmentation \cite{badrinarayanan2017segnet}, etc. It is, therefore, common to expect that successful deep models can obtain good performances. However, it has been shown that, in practice, these approaches easily fail to generalize well \cite{zhang2016understanding}. 

According to the literature, the most important reasons for this failure are (i) the small size of the training data which causes overfitting and (ii) the large difference between training and test data which is typically addressed as domain shift. Therefore, one of the most important problems that arises is how to improve the quality of models so that they generalize well to unseen data from a different domain. During the last years, several algorithms have been proposed to tackle the mentioned problem, improving the  models generalization through heuristic techniques such as dropout \cite{srivastava2014dropout}, early stopping \cite{montavon2012neural}, weight decay \cite{krogh1992simple}, data augmentation \cite{Jaderberg:2015:STN:2969442.2969465}, and randomization methods \cite{zhang2016understanding}. 

Thanks to these regularization methods, deep models are reaching expert-level accuracy in medical image segmentation. However, they still have a limited clinical application due to the aforementioned challenge (i), which is also considered as one of the most relevant and common problem in medical image analysis tasks \cite{carass2017longitudinal}. To tackle this challenge, several strategies have been proposed, such as using 2.5D (slices) rather than full-size 3D images \cite{aslani2019multi}, initializing parameters of the proposed model with pre-trained weights on natural images \cite{aslani2018deep}, and adopting special data augmentation techniques \cite{zhang2019unseen}. An effective solution, however, would be to merge datasets collected from different centers. This, however, introduces another important challenge. The medical data acquisition can vary significantly between different centers. For instance, in magnetic resonance imaging (MRI), this procedure is often subject to the variation of several specific properties such as scanner, magnet strength, and acquisition protocol. This causes high domain variability between datasets which eventually can result in poor generalization. In order to tackle this problem, several methods have been proposed such as scanner invariant representations for medical image harmonization \cite{moyer2019scanner}, one-shot domain adaptation \cite{valverde2019one}, and unsupervised methods \cite{perone2019unsupervised} for medical image segmentation.

In this work, we propose a novel simple generalization method to enhance baseline models and diminish the effect of domain differences in the data. To this end, a regularization network, equipped with an auxiliary loss function, is proposed to incorporate regularization into a standard encoder-decoder segmentation network. We tested the model with a standard cross-validation procedure using an in-house dataset on Multiple Sclerosis (MS) lesion segmentation. Results show that the proposed regularization network has a significant impact on the generalization of the standard segmentation network when data from multiple centers are used.

\begin{figure*}[t!]
\centering
\includegraphics[width=99mm]{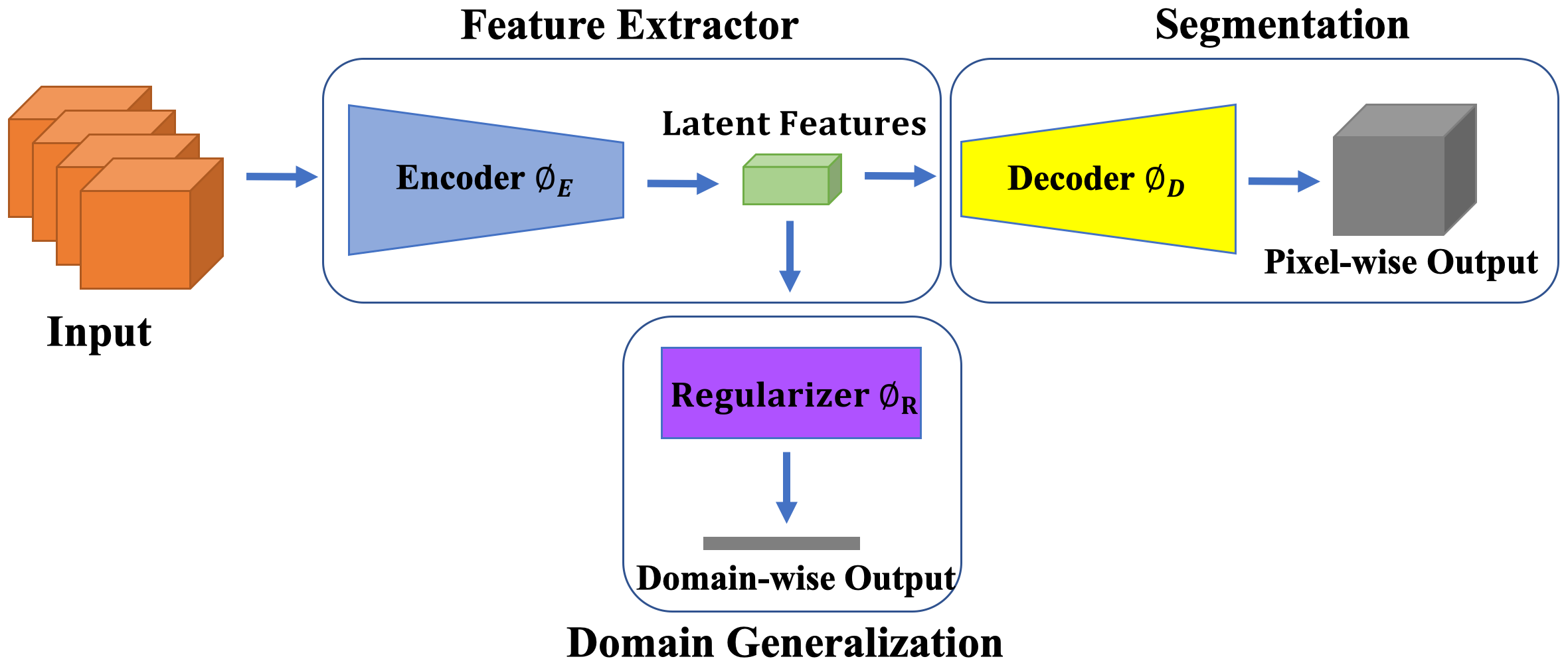}
\caption{Overall architecture of the proposed method including three main components: An encoder network $\phi_E$ for extracting latent features, a decoder network $\phi_D$ for segmentation and a regularization network $\phi_R$ for domain generalization.}  
\label{fig1}
\end{figure*}

\section{Method}
\label{method}
Generally, the performance of models suffers when they are applied to domains other than the ones they were trained upon.In this work, the goal is to improve the generalizability of a backbone segmentation model by training on a collection of datasets presenting high domain variability. The selected backbone encoder-decoder model has a traditional loss function used for image segmentation. However, to handle the domain shift problem, we propose a regularization network including an auxiliary loss function that is designed to encourage the model to ignore domain-specific information. This property emerges from optimizing cross entropy or correlation coefficient as detailed below. Training the backbone segmentation network incorporated with regularization network reduces the domain differences problem across the datasets.

\subsection{Network Architecture}
\label{network architecture}
The overall architecture of the proposed model includes three main components (\autoref{fig1}). The first component is a feature extractor network consisting of an encoder $\phi_E$ which is fed by an input image $x_i \in \mathcal{X}$. The output of the encoder $\phi_E$ is a $p$-dimensional vector  $r_i=\{ r_{ij} \in \mathbb{R} \}_{j=1}^p\in \mathcal{R}$ representing the latent features. 

The second component is a segmentation network consisting of a decoder $\phi_D$ which reconstruct from the latent features $r_i\in\mathcal{R}$ a feature representation with the resolution of the input image $x_i \in \mathcal{X}$. The output layer then produces a dense pixel-wise prediction output $s_i\in\mathcal{S}$ using a softmax activation. This network includes a traditional loss term used to update the weights to improve the segmentation performance (see Section~\ref{sec:loss functions}). Note that the described architecture composed of the mentioned encoder $\phi_E$ and decoder $\phi_D$ is very similar to the model presented in \cite{ronneberger2015u}, using 3D operations rather than 2D ones and removing the regularization layers (dropout). For simplicity, we removed skip-connections in \autoref{fig1}. 

The third component of the presented model is a regularization network $\phi_R$ including three perceptron layers and a softmax layer. The network receives the latent features $r_i\in\mathcal{R}$ to produce category-wise prediction $c_i\in\mathcal{C}$, which in our case corresponds to the prediction of the input's domain. Our observation is that during training, the model without $\phi_R$ learns how to segment the input images also encoding their source domain. This results in overfitting of the model with a domain bias. As a result, the segmentation performance of the network on data coming from unseen source domains degrades. The goal of the regularization network is therefore to steer the whole model to reduce the domain bias, to obtain a better generalization and, hence, a fairer segmentation performance on seen and unseen domains. To this aim, we introduce an auxiliary loss term whose aim is to confuse the model about the dataset domains, thus forcing the model to learn how to segment the image while maximally reducing the domain bias.

\subsection{The Loss Functions}
\label{sec:loss functions}
Our method has been tested using multiple loss terms to enable the network to precisely segment the input image while generalizing over the domains. Specifically, the proposed model was optimized according to the loss function formulated as:

\begin{equation}
\begin{split}
\mathcal{L}(\mathcal{X,S,C,H,G})= \mathcal{L}_{\rm seg}(\mathcal{X,S,G}) + \\
\lambda\mathcal{L}_{\rm reg}(\mathcal{X,C,H})
\label{full_loss}
\end{split}
\end{equation}
 
\noindent where $\lambda \in [0,1]$ is a hyperparameter controlling the trade-off between segmentation and regularization losses. $\mathcal{H}$ and $\mathcal{G}$ are the domain-wise and pixel-wise ground truth, respectively. In this work, we propose three different regularization approaches using loss functions $\mathcal{L}_{\rm reg}$ based on two well-established measures, namely cross-entropy and the Pearson correlation coefficients. Hence, $\mathcal{L}_{\rm reg}$ can  take  on either of these three options:
 
\begin{equation}
\begin{split}
\mathcal{L}_{\rm reg}(\mathcal{X,C,H})= 
\{\mathcal{L}_{\rm pc}(\mathcal{X,C,H}),&  
\mathcal{L}_{\rm rand}(\mathcal{X,C,H}), \\ 
\mathcal{L}_{\rm du}(\mathcal{X,C,H})\}
\end{split}
\end{equation}

\noindent\textbf{Pearson Correlation Loss $\mathcal{L}_{\rm pc}$:}
Given an input image $x_i\in\mathcal{X}$, the regularization network $\phi_R$ generates the corresponding output vector $c_i=\{ c_{ij} \in [0,1]\}_{j=1}^n\in\mathcal{C}$ which shows the probabilities for $x_i$ to be in one of the $n$ domains. For each input image $x_i$ the corresponding one-hot encoded vector as ground truth domain labeling is also given by $h_i = (0, 0,...,1,0,...,0)\in\mathcal{H}$. 

The Pearson correlation coefficient measures the strength of linear correlation or similarity between two variables, where higher values correspond to higher similarity. Hence, to remove the domain bias, the model can be trained to minimize the Pearson correlation between $\mathcal{C}$ and $\mathcal{H}$

\begin{equation}
\mathcal{L}_{\rm pc}(x_i,c_i,h_i)= \frac{\sum_j(c_{ij}-\overline{c_i})(h_{ij}-\overline{h_i})}{\sqrt{\sum_j(c_{ij}-\overline{c_i})^2 \sum_j(h_{ij}-\overline{h_i})^2}}
\end{equation}

\noindent where $\overline{h_i}$ and $\overline{c_i}$ denote the mean values of elements in the vectors $h_i$ and $c_i$, respectively.\\

\begin{figure*}[t!]
\centering
\includegraphics[width=165mm]{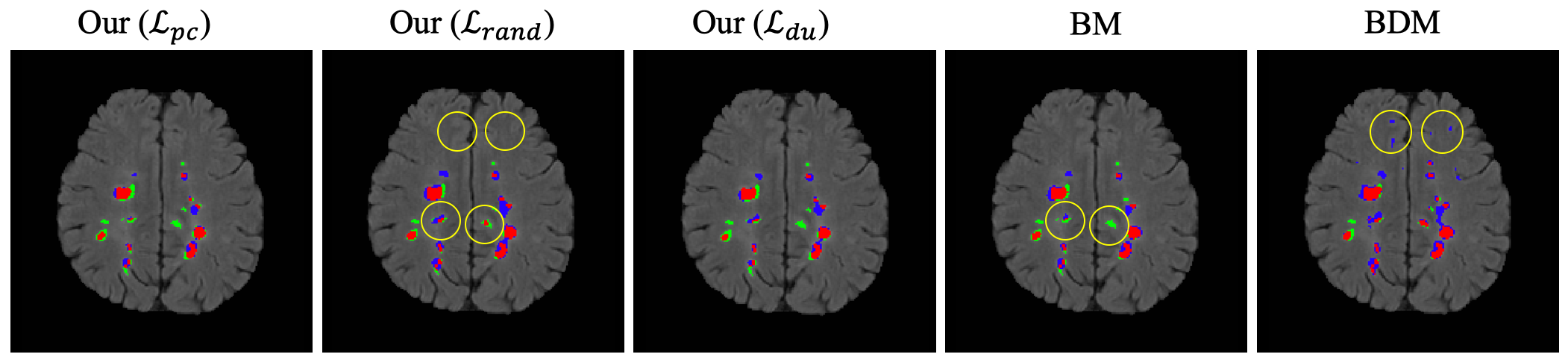}
\caption{Segmentation of a random subject obtained by different methods against ground truth annotation. On all images, true positives, false negatives, and false positives are marked in red, green and blue, respectively (refer to Section~\ref{sec: disc} for yellow circles).}  
\label{fig2}
\end{figure*}

\noindent\textbf{Randomized Cross-Entropy Loss $\mathcal{L}_{\rm rand}$:}
The most commonly used loss function for image classification is the cross-entropy:

\begin{equation}
\mathcal{L}_{\rm rand}(x_i,c_i,h_i)=-\sum_j h_{ij}\log c_{ij}
\label{eq:cross-entropy}
\end{equation}

\noindent which allows comparing the class predictions vector $c_i\in\mathcal{C}$ and the ground truth one-hot encoded vector $h_i\in\mathcal{H}$, penalizing the correct classes having a probability diverging from the expected value. Our ultimate goal is to push the network to filter unnecessary domain information during training. This can be easily obtained by shuffling the ground truth $h_i\in\mathcal{H}$ at each training iteration and for every single input. This encourages the model to learn the wrong classes selected randomly.\\

\noindent\textbf{Discrete Uniform Loss $\mathcal{L}_{\rm du}$:} Analyzing the problem from a different prospective, to remove the domain bias, the encoder network $\phi_E$ should generate a representation $r_i\in\mathcal{R}$ from which the domain classifier in $\phi_R$ cannot extract information. This should correspond to a classifier in $\phi_R$ that classify any class with equal probability, independently from the input. We can obtain this result training the model with the cross-entropy loss (\autoref{eq:cross-entropy}), forcing the domain ground truth to be a uniform distribution
$h_i: \{ h_{ij}=\frac{1}{n}\}_{j=1}^n\in \mathcal{H}$\\

\noindent\textbf{Segmentation Loss $\mathcal{L}_{\rm seg}$:}
To fit the model according to the segmentation task, we used a well-known loss function for image segmentation, namely the soft-Dice loss function:

\begin{equation}
\label{eq:DL}
\mathcal{L}_{seg}(x_i,s_i,g_i) =1-\frac{2 \sum_i s_i g_i} {\sum_i s_i^2 + \sum_i g_i^2}
\end{equation}

\noindent where  $s_i\in\mathcal{S}$ is the dense pixel-wise prediction and $g_i\in\mathcal{G}$ is the corresponding  ground truth segmentation. This function penalizes $\phi_E$ and $\phi_D$ based on the overlap between the prediction and the ground truth segmentation.

\subsection{Implementation Details}
We designed a backbone model based on the U-Net \cite{ronneberger2015u}, built by concatenating a down-sampling encoder $\phi_E$ made by 4 stages with an up-sampling decoder $\phi_D$ made by 4 stages. However, differently from the U-Net, the regularization layers (dropout) were removed and all 2D operations were replaced by their 3D counterparts. During the training procedure, the backbone model was combined with the proposed regularization network. On the contrary, during the testing phase, the regularization network was removed addressing only the segmentation task. 

The model was executed on 3D patches with size $64\times64\times64$ cropped from each volume (with 50\% overlap). The model was trained only using patches containing at least one voxel labeled as a lesion. During the test, on the contrary, all patches were used. The evaluation of the model was performed on the reconstructed full-size volumes, fusing the predictions for all patches. The proposed model was implemented in Python using Keras with Tensorflow backend. We trained our model using Adam optimizer with an initial learning rate of 0.0001. The size of batch and the maximum number of training epochs were fixed respectively at 15 and 500 (with 300 steps per epoch). Regarding the model initialization, all blocks were randomly initialized from a Gaussian distribution. The hyperparameter $\lambda$ in \autoref{full_loss} was selected through grid search with values equal to 0.2, 0.3 and 0.1 for $\mathcal{L}_{\rm pc}$, $\mathcal{L}_{\rm du}$ and $\mathcal{L}_{\rm rand}$, respectively.

\section{Experiments}
\label{experiment}
We evaluate the performance of the proposed method on an in-house clinical dataset, collected from 56 different centers with a variable number of MS patients per center. Each patient had several scanning sessions, with each session including 4 MRI modalities (T1w, T2w, PDw, and FLAIR). Each volume is composed by 60 slices with FOV=256$\times$256 and $1mm\times1mm\times3mm$ voxel resolution. All volumes were already segmented manually representing the ground truth lesion masks. 

All images were skull-stripped using Brain Extraction Tool (BET) \cite{smith2002fast}, and rigidly registered to the $1mm^3$ MNI-ICBM152 template \cite{oishi2008human} using FMRIB's Linear Image Registration tool (FLIRT) \cite{jenkinson2002improved}. Ethics approval for data collection and secondary analysis was granted by a Research Ethics Board at the University of British Columbia.\\

\noindent\textbf{Experimental Protocol:}
We considered each site as a separate domain and to keep data balanced over all available sites, a single subject including one time point with four MRI modalities (T1w, T2w, PDw, and FLAIR) was selected from each site. We implemented 5-fold cross-validation over the whole data (60\%:20\%:20\% for training, validation, and test, respectively). For comparison purpose, we repeated the above-mentioned experiment (using exactly the same folds) for the backbone model without any regularization (denoted as the BM) and the same backbone model with additional dropout layers (denoted as the BDM). Note that BDM model is equivalent the U-net model \cite{ronneberger2015u} with replacing the 2D operations by their 3D counterparts.

\begin{figure}[t!]
\centering
\includegraphics[width=80mm]{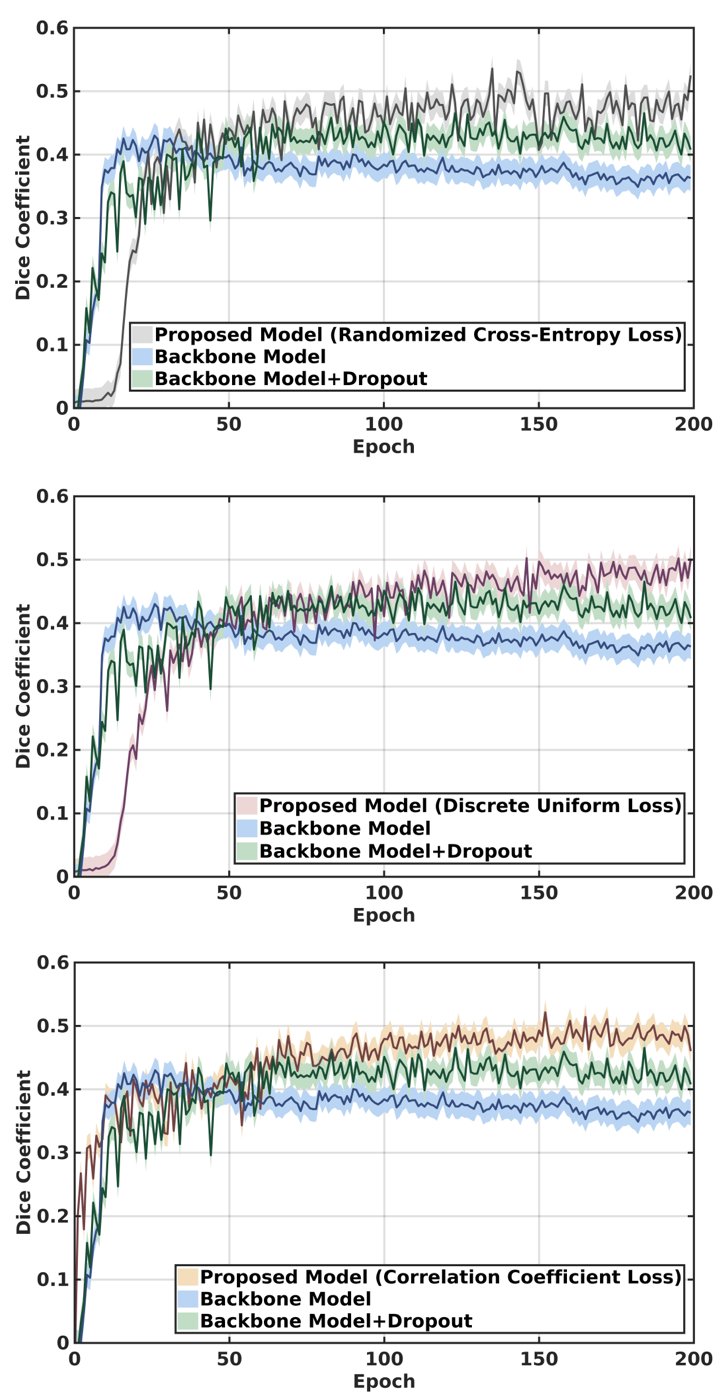}
\caption{Comparison of the DSC measure performance of the proposed methods with other baseline models on validation set during training.} 
\label{fig3}
\end{figure}

\section{Results}
We evaluated our model using four measures: Dice Similarity Coefficient (DSC), Lesion-wise True Positive Rate (LTPR), Lesion-wise False Positive Rate (LFPR), and Positive Prediction Value (PPV) (refer to \cite{carass2017longitudinal} for more details). \autoref{table1} summarizes the results of our experiment on the test set comparing our model with other baseline methods. The Table shows the mean value of DSC, LTPR, LFPR, and PPV. As can be seen, our proposed methods outperformed baseline methods on the DSC measure. Moreover, in terms of LTPR and LFPR measures, our  model with the randomized and discrete uniform auxiliary loss functions showed more balanced performance compared with the other models \autoref{fig2} shows an example of segmentation for all methods. \autoref{fig3} compares the DSC performance of the proposed methods with other models on the validation set. Confirming the results reported in the test set, as shown in this Figure, our model with all three possible auxiliary loss terms depicts better DSC performance than the baseline methods. 

\begin{table}[t!]
\scriptsize
\centering
\caption{Results related to our experiment. Mean values of DSC, LTPR, LFPR, and PPV were measured for different methods. Values in bold and italic indicate the first-best and second-best results.}
\scalebox{1.28}{
\begin{tabular}{c|cccc}
\hline
Method & DSC & LTPR & LFPR & PPV \\  \hline
Our $(\mathcal{L}_{\rm pc})$ & 0.4638 & 0.4267 & 0.3954 & 0.4865 \\
Our $(\mathcal{L}_{\rm rand})$ & \textbf{0.5001} & 0.4618 & \textbf{0.3348} & \textbf{0.5193} \\
Our $(\mathcal{L}_{\rm du})$ & \textit{0.4893} & \textit{0.4670} & 0.3525 & \textit{0.5182} \\
BM & 0.4540 & 0.4318 & \textit{0.3383} & 0.5088 \\
BDM & 0.4598 & \textbf{0.5821} & 0.5151 & 0.4577 \\
\hline
\end{tabular}
}
\label{table1}
\end{table}

\section{Discussion and Conclusion}
\label{sec: disc}
In this paper, we have introduced a generalization methods implemented via an auxiliary loss with three variants. We tested this method on medical image segmentation, particularly MS lesion segmentation from MRI modalities in the presence of domain shift originating from multi-center dataset. The proposed model is the combination of a traditional encoder-decoder network for segmentation and an additional regularization network including an auxiliary loss term for domain generalization.
 
Investigating the impact of the proposed method summarized in \autoref{table1}, our model always outperformed the baseline models when considering the DSC measures (regardless of which of the adopted auxiliary loss variant was used). However, the best performance in terms of DSC, LFPR, and PPV measures among all tested models is provided by our model with the randomized auxiliary loss function. The BDM model showed the best LTPR measure together with the worst LFPR measure showing that this model has very poor trade-off between LTPR and LFPR. Confirming the above-mentioned point, \autoref{fig2} shows that BDM model has over-segmented lesion regions (referring to the top two yellow circles). Moreover, it can be seen that BM model did not identify some lesions (referring to the bottom two yellow circles). However, the proposed method with randomized loss term shows a considerable good performance by not only identifying the mentioned small lesions but also ignoring the false positives.\\

\noindent\textbf{Acknowledgments.} We gratefully acknowledge Medical Image Analysis Lab members for their helpful comments and Compute Canada for computational resources.

{\small
\bibliographystyle{ieeetr}
\bibliography{egbib}

\begin{thebibliography}{10}

\bibitem{lecun1998gradient}
Y.~LeCun, L.~Bottou, Y.~Bengio, P.~Haffner, {\em et~al.}, ``Gradient-based
  learning applied to document recognition,'' {\em Proceedings of the IEEE},
  vol.~86, no.~11, pp.~2278--2324, 1998.

\bibitem{huang2017densely}
G.~Huang, Z.~Liu, L.~Van Der~Maaten, and K.~Q. Weinberger, ``Densely connected
  convolutional networks,'' in {\em Proceedings of the IEEE {Conference} on
  {Computer} {Vision} and {Pattern} {Recognition}}, pp.~4700--4708, 2017.

\bibitem{girshick2014rich}
R.~Girshick, J.~Donahue, T.~Darrell, and J.~Malik, ``Rich feature hierarchies
  for accurate object detection and semantic segmentation,'' in {\em
  Proceedings of the IEEE {Conference} on {Computer} {Vision} and {Pattern}
  {Recognition}}, pp.~580--587, 2014.

\bibitem{badrinarayanan2017segnet}
V.~Badrinarayanan, A.~Kendall, and R.~Cipolla, ``Segnet: A deep convolutional
  encoder-decoder architecture for image segmentation,'' {\em IEEE
  {Transactions} on {Pattern} {Analysis} and {Machine} {Intelligence}},
  vol.~39, no.~12, pp.~2481--2495, 2017.

\bibitem{zhang2016understanding}
C.~Zhang, S.~Bengio, M.~Hardt, B.~Recht, and O.~Vinyals, ``Understanding deep
  learning requires rethinking generalization,'' {\em arXiv preprint
  arXiv:1611.03530}, 2016.

\bibitem{srivastava2014dropout}
N.~Srivastava, G.~Hinton, A.~Krizhevsky, I.~Sutskever, and R.~Salakhutdinov,
  ``Dropout: a simple way to prevent neural networks from overfitting,'' {\em
  The {Journal} of {Machine} {Learning} {Research}}, vol.~15, no.~1,
  pp.~1929--1958, 2014.

\bibitem{montavon2012neural}
G.~Montavon, G.~Orr, and K.-R. M{\"u}ller, {\em Neural networks: tricks of the
  trade}, vol.~7700.
\newblock springer, 2012.

\bibitem{krogh1992simple}
A.~Krogh and J.~A. Hertz, ``A simple weight decay can improve generalization,''
  in {\em Advances in {Neural} {Information} {Processing} {Systems}},
  pp.~950--957, 1992.

\bibitem{Jaderberg:2015:STN:2969442.2969465}
M.~Jaderberg, K.~Simonyan, A.~Zisserman, and K.~Kavukcuoglu, ``Spatial
  transformer networks,'' in {\em Proceedings of the 28th International
  Conference on Neural Information Processing Systems - Volume 2}, NIPS'15,
  (Cambridge, MA, USA), pp.~2017--2025, MIT Press, 2015.

\bibitem{carass2017longitudinal}
A.~Carass, S.~Roy, A.~Jog, J.~L. Cuzzocreo, E.~Magrath, A.~Gherman, J.~Button,
  J.~Nguyen, F.~Prados, C.~H. Sudre, {\em et~al.}, ``Longitudinal multiple
  sclerosis lesion segmentation: Resource and challenge,'' {\em NeuroImage},
  vol.~148, pp.~77--102, 2017.

\bibitem{aslani2019multi}
S.~Aslani, M.~Dayan, L.~Storelli, M.~Filippi, V.~Murino, M.~A. Rocca, and
  D.~Sona, ``Multi-branch convolutional neural network for multiple sclerosis
  lesion segmentation,'' {\em NeuroImage}, vol.~196, pp.~1--15, 2019.

\bibitem{aslani2018deep}
S.~Aslani, M.~Dayan, V.~Murino, and D.~Sona, ``Deep {2D} encoder-decoder
  convolutional neural network for multiple sclerosis lesion segmentation in
  brain mri,'' in {\em International MICCAI Brainlesion Workshop},
  pp.~132--141, Springer, 2018.

\bibitem{zhang2019unseen}
L.~Zhang, X.~Wang, D.~Yang, T.~Sanford, S.~Harmon, B.~Turkbey, H.~Roth,
  A.~Myronenko, D.~Xu, and Z.~Xu, ``When unseen domain generalization is
  unnecessary? rethinking data augmentation,'' {\em arXiv preprint
  arXiv:1906.03347}, 2019.

\bibitem{moyer2019scanner}
D.~Moyer, G.~V. Steeg, C.~M. Tax, and P.~M. Thompson, ``Scanner invariant
  representations for diffusion mri harmonization,'' {\em arXiv preprint
  arXiv:1904.05375}, 2019.

\bibitem{valverde2019one}
S.~Valverde, M.~Salem, M.~Cabezas, D.~Pareto, J.~C. Vilanova,
  L.~Rami{\'o}-Torrent{\`a}, {\`A}.~Rovira, J.~Salvi, A.~Oliver, and
  X.~Llad{\'o}, ``One-shot domain adaptation in multiple sclerosis lesion
  segmentation using convolutional neural networks,'' {\em NeuroImage:
  Clinical}, vol.~21, p.~101638, 2019.

\bibitem{perone2019unsupervised}
C.~S. Perone, P.~Ballester, R.~C. Barros, and J.~Cohen-Adad, ``Unsupervised
  domain adaptation for medical imaging segmentation with self-ensembling,''
  {\em NeuroImage}, vol.~194, pp.~1--11, 2019.

\bibitem{ronneberger2015u}
O.~Ronneberger, P.~Fischer, and T.~Brox, ``U-net: Convolutional networks for
  biomedical image segmentation,'' in {\em International Conference on Medical
  {Image} {Computing} and {Computer-assisted} {Intervention}}, pp.~234--241,
  Springer, 2015.

\bibitem{smith2002fast}
S.~M. Smith, ``Fast robust automated brain extraction,'' {\em Human {Brain}
  {Mapping}}, vol.~17, no.~3, pp.~143--155, 2002.

\bibitem{oishi2008human}
K.~Oishi, K.~Zilles, K.~Amunts, A.~Faria, H.~Jiang, X.~Li, K.~Akhter, K.~Hua,
  R.~Woods, A.~W. Toga, {\em et~al.}, ``Human brain white matter atlas:
  identification and assignment of common anatomical structures in superficial
  white matter,'' {\em Neuroimage}, vol.~43, no.~3, pp.~447--457, 2008.

\bibitem{jenkinson2002improved}
M.~Jenkinson, P.~Bannister, M.~Brady, and S.~Smith, ``Improved optimization for
  the robust and accurate linear registration and motion correction of brain
  images,'' {\em Neuroimage}, vol.~17, no.~2, pp.~825--841, 2002.

\end{thebibliography}
}
\end{document}